\documentclass[10pt,twocolumn,notitlepage,sort&compress]{article}
\usepackage [english]{babel}
\usepackage[utf8]{inputenc}

\usepackage{graphicx} 
\usepackage{dcolumn} 
\usepackage{bm} 
\usepackage{amssymb}
\usepackage{amsmath}
\usepackage{subcaption}
\usepackage{xcolor}
\usepackage[margin=0.5in]{geometry}
\usepackage{caption}
\usepackage{setspace}
\usepackage{multicol}
\usepackage{authblk}
\usepackage{xfrac}
\usepackage{multirow,tabularx}
\usepackage{soul}

\usepackage[resetlabels,labeled]{multibib}
\newcites{Supp}{References}

\newcommand{\ME}{{\mathcal{E}}}
\newcommand{\olE}{{\overline{E}}}

{\vskip 2pt \begin{compactitem}[#1]\setlength{\itemsep}{2pt}}
{\end{compactitem}\vskip 2pt}

\newcommand{\be}{\begin{equation}}
\newcommand{\ee}{\end{equation}} 
\newcommand{\lb}{\label}
\newcommand{\OL}{\overline}

\newcommand{\br}{{\bf r}}
\newcommand{\bu}{{\bf u}}

\newcommand{\bx}{{\bf x}}

\newcommand{\cO}{{\mathcal O}}

\begin{document}



\title{Global Energy Spectrum of the General Oceanic Circulation}

\author[1]{Benjamin A. Storer}
\author[2]{Michele Buzzicotti}
\author[3]{Hemant Khatri}
\author[4]{Stephen M. Griffies}
\author[1]{Hussein Aluie\thanks{hussein@rochester.edu}}


\affil[1]{Department of Mechanical Engineering and Laboratory for Laser Energetics, University of Rochester, Rochester, New York, USA}

\affil[2]{Department of Physics, University of Rome Tor Vergata and INFN, Rome, Italy}

\affil[3]{Department of Earth, Ocean, and Ecological Sciences, University of Liverpool, UK}

\affil[4]{NOAA Geophysical Fluid Dynamics Laboratory and Princeton University Atmospheric and Oceanic Sciences Program, Princeton, New Jersey, USA}

\date{\today}

\maketitle

{\bf 
Since the advent of satellite altimetry, our perception of the oceanic circulation has brought into focus the pervasiveness of mesoscale eddies that have typical scales of tens to hundreds of kilometers \cite{chelton2011global}, are the ocean's analogue of weather systems, and are often thought of as the peak of the ocean's kinetic energy (KE) wavenumber spectrum \cite{FerrariWunsch09,sasaki2014impact,Torresetal2018jgr}. 
Yet, our understanding of the ocean’s spatial scales has been derived mostly from Fourier analysis in small ``representative'' regions (e.g. \cite{Qiuetal2018jpo,ORourke_etal2018,CalliesWu2019}), typically a few hundred kilometers in size, that cannot capture the vast dynamic range at planetary scales. 
Here, we present the first truly global wavenumber spectrum of the oceanic circulation from satellite data and high-resolution re-analysis data, using a coarse-graining method to analyze scales much larger than what had been possible before. 
Spectra spanning over three orders of magnitude in length-scale reveal the Antarctic Circumpolar Current (ACC) as the spectral peak of the global extra-tropical ocean, at $\approx10\times10^3~$km. 
We also find a previously unobserved power-law scaling over scales larger than $10^3~$km.
A smaller spectral peak exists at $\approx300~$km associated with the mesoscales, which, due to their wider spread in wavenumber space, account for more than $50\%$ of the resolved surface KE globally. 
Length-scales that are twice as large (up to \(10^3\)~km) exhibit a characteristic lag time of \(\approx40~\)days in their seasonal cycle, such that in both hemispheres KE at $100~$km peaks in late spring while KE at $10^3~$km peaks in late summer. 
The spectrum presented here affords us a new window for understanding the multiscale general oceanic circulation within Earth's climate system, including the largest planetary scales.
}

The oceanic circulation is a key component in Earth's climate system.
It is both the manifestation and cause of a suite of linear and nonlinear dynamical processes acting over a broad range of scales in both space and time \cite{FerrariWunsch09}. 
The wavenumber spectrum of the oceanic circulation allows us to understand the energy distribution across spatial scales throughout the globe, reveals key bands of scales within the circulation system at which energy is concentrated, and unravels power-law scalings that can be compared to theoretical predictions \cite{Vallis17}. 
The spectrum is an important guide to probing (i) energy sources and sinks maintaining the oceanic circulation at various scales, (ii) how energy is ultimately dissipated, and (iii) how the ocean at a global climate scale is coupled to motions several orders of magnitude smaller.

Thanks to satellite observations \cite{Kleinetal2019} and high-resolution models and analysis \cite{MITGCM48deg, Qiuetal2018jpo}, it is now well-appreciated that the mesoscales, traditionally thought of as transient eddies of $\cO(100)~$km in size, form a key band of spatial scales that pervade the entire ocean and have a leading order effect on the transport of heat, salt, and nutrients, as well as coupling to the global meridional overturning circulation \cite{marshall2017dependence}. 
The mesoscales are generally viewed as forming the peak of the KE spectrum of the oceanic circulation \cite{FerrariWunsch09,Kleinetal2019} 
(e.g. Fig. 5 in \cite{sasaki2014impact} or Fig. 5 in \cite{Torresetal2018jgr}). 
However, the existence of the mesoscale spectral peak and the length-scale at which it occurs is not known with certainty \cite{FerrariWunsch09}. 
Evidence is often derived from performing Fourier analysis on the ocean surface velocity \cite{Qiuetal2018jpo} or sea-surface height \cite{ORourke_etal2018} within regions that are typically 5$^\circ$ to 10$^\circ$ in extent (nominally $500~$km to $10^3~$km) \cite{Stammer1997ty}. 
The peak appears in only a fraction of the chosen regions, and spectral energy tends to be largest at the largest length-scales (smallest wavenumbers), which are most susceptible to artifacts from the finite size of the chosen regions and the windowing required for Fourier analysis \cite{FerrariWunsch09}. 
To date, there has been no determination of the oceanic energy wavenumber spectrum at planetary scales. 
Do the mesoscales of $\cO(100)~$km actually form the peak of the ocean's KE spectrum? 
What is the KE content of scales larger than $\cO(10^3)~$km, which constitute the ocean's gyres and are directly coupled to the climate system?

Below, we present the first KE spectrum over the entire range of scales resolved in data from satellites and high-resolution models at the ocean's surface, including the spectrum at planetary scales. 
We find that the spectral peak of the global extratropical ocean is at ${\approx10\times10^3~}$km and is due to the ACC. 
We see vestiges of a similar peak in the northern hemisphere, which is arrested at a smaller amplitude and at smaller scales ($\approx 4\times10^3~$km) due to continental boundaries. 
Another prominent spectral peak is at $\approx 300~$km, and with an amplitude less than half that of the ACC. 
Yet, the cumulative energy in the mesoscales between $100~$km and $500~$km is very large ($>50\%$ of total resolved energy). 
We also report the first observation of a roughly $k^{-1}$ power-law scaling over scales larger than $10^3~$km in both hemispheres, consistent with a theoretical prediction from a quasigeostrophic model forced by wind \cite{MullerFrankignoul1981,Wortham2014}, with the power-law scaling extending up to the ACC peak in the southern hemisphere. 

Our results here open exciting avenues of inquiry into oceanic dynamics, allowing us to seamlessly probe interactions between motions at scales $\cO(100)~$km and smaller with planetary scales larger than $\mathcal{O}(10^3)~$km relevant to climate. 
We are able to do so using a coarse-graining approach developed recently to probe multi-scale geophysical processes \cite{aluie2018mapping,aluie2019convolutions,Raietal2021}.

\subsection*{Partitioning energy across length-scales}

Our methodology, described in the Methods section and in \cite{aluie2018mapping,aluie2019convolutions}, allows us to coarse-grain the ocean flow at any length-scale of choice and calculate the KE of the resulting coarse flow. 
By performing a `scan' over an entire range of length-scales, we extract the so-called `filtering spectrum' without needing to perform Fourier transforms \cite{sadek2018extracting}. 
The filtering spectrum and the traditional Fourier spectrum agree when the latter is possible to calculate, as demonstrated in \cite{sadek2018extracting} and in Fig.~\ref{fig:Qiu_comparison} of the Methods section. 
Unlike traditional Fourier analysis within a box / subdomain, coarse-graining can be meaningfully applied on the \textit{entire} spherical planet, including land/sea boundaries, and so allows us to probe everything from the smallest resolved scales up to true planetary scales.

\begin{figure*}[ht]
    \centering
    \includegraphics[scale = 1]{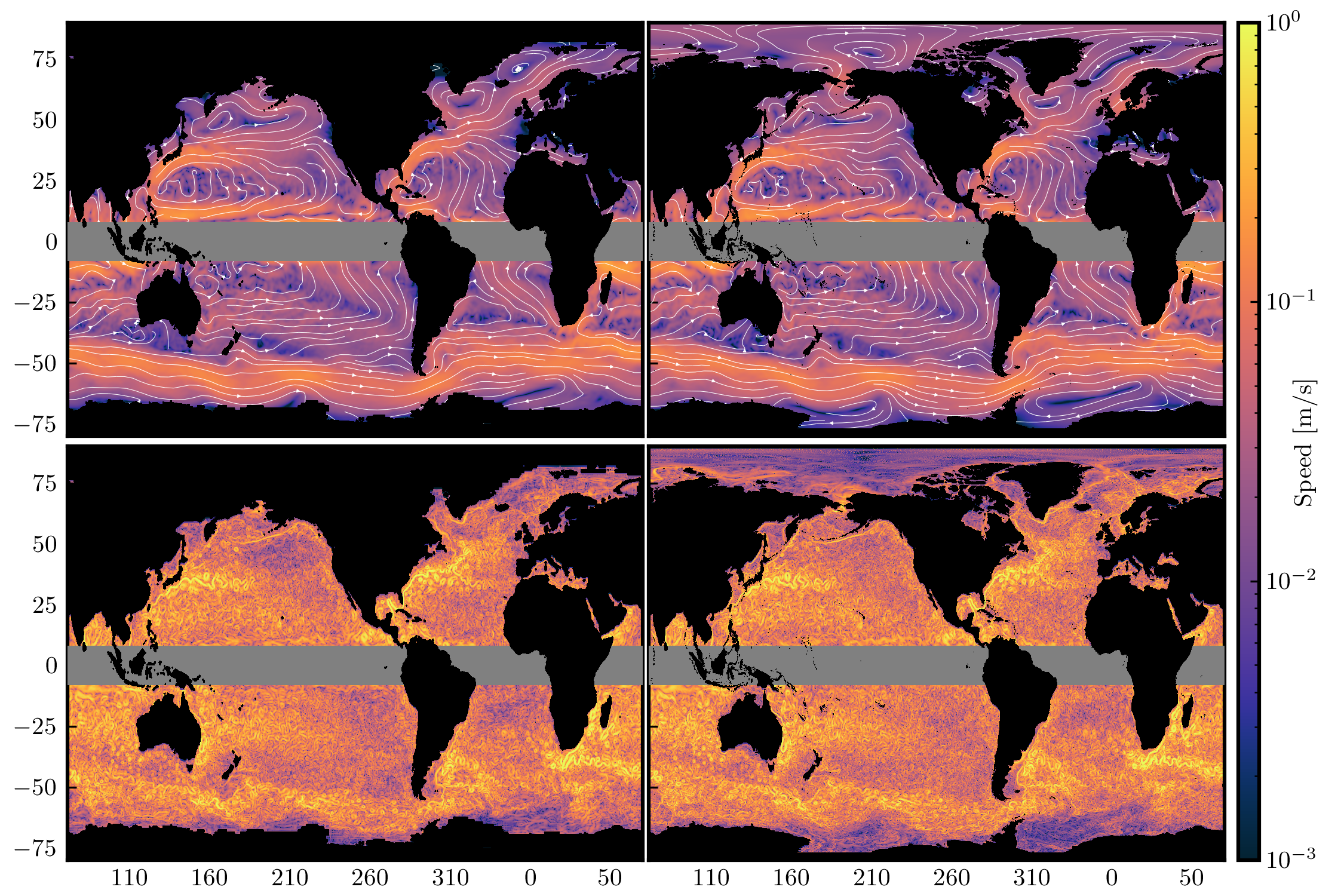}
    \caption{
        \textbf{ Gyre-scale and mesoscale flows }
        \textbf{ [Colour maps]} show the geostrophic velocity magnitude for length scales \textbf{[top]} larger than \(10^3~\)km and \textbf{[bottom]} smaller than \(10^3~\)km for a single day (02 Jan 2015).
        \textbf{[Left]} shows the \(\sfrac{1}{4}^\circ\) AVISO dataset
        and \textbf{[right]} the \(\sfrac{1}{12}^\circ\) NEMO dataset. 
        \textbf{[White lines]} highlight the corresponding streamlines, with arrows showing the direction of the flow. Areas in black include land, and also ice coverage in the AVISO dataset. 
        In this work we exclude the tropics where velocity from satellite altimetry is less reliable, and define the northern and southern hemispheres (NH and SH, respectively) as the ocean poleward of \(15^\circ\).
    }
    \label{fig:maps:NSH_streamlines_AVISO_NEMO}
\end{figure*}

\paragraph*{ Filtering Spectrum }
Given a velocity field \(\mathbf{u}\) and a filter scale \(\ell\), coarse-graining produces a filtered velocity $\OL{\bu}_\ell$ that only contains spatial scales larger than $\ell$, having had smaller scales removed (see Fig.~\ref{fig:maps:NSH_streamlines_AVISO_NEMO} and the Methods section). 
Unlike standard approaches to low-pass filtering geophysical flows, such as by averaging adjacent grid-cells or block-averaging in latitude-longitude, the coarse-graining of \cite{aluie2019convolutions} used here relies on a generalized convolution operation that respects the underlying spherical topology of the planet, thus preserving the fundamental physical properties of the flow, such as its incompressibility, its geostrophic character, and the vorticity present at various scales.
The KE (per unit mass, in m$^2$/s$^2$) contained in scales larger than $\ell$ is 
\be
\ME_\ell = \frac{1}{2} \left|\OL{\bu}_\ell(\bx,t)\right|^2 \hspace*{0.5cm}\mbox{\text{(coarse KE).}}
\label{eq:coar_kene}
\ee
While $\ME_\ell$ quantifies the \emph{cumulative} energy at {all} scales larger than $\ell$, the wavenumber spectrum quantifies the spectral energy \emph{density} at a \emph{specific} scale, similar to the common Fourier spectrum.
Following \cite{sadek2018extracting}, we extract the KE content at different length-scales by differentiating in scale the coarse KE:
\be
\olE(k_\ell,t) = \frac{d}{dk_\ell} \left \{ \ME_{\ell} \right \} = -\ell^2 \frac{d}{d \ell} \left \{ \ME_\ell \right \},
\label{eq:filt_spect}
\ee
where $k_\ell = 1/\ell$ is the `filtering wavenumber' and $\{\cdot\}$ denotes a spatial average. 
Ref \cite{sadek2018extracting} identified the conditions on the coarse-graining kernel for $\olE(k_\ell,t)$ to be meaningful in the sense that its scaling agrees with that of the traditional Fourier spectrum when Fourier analysis is possible, such as in periodic domains. 
Fig.~\ref{fig:Qiu_comparison} in Methods shows how the filtering spectrum agrees with the Fourier spectrum performed within an oceanic box region over length-scales smaller than the box but has the important advantage of quantifying larger scales without being artificially limited by the box size and windowing functions to synthetically periodize the data.

\paragraph*{Oceanic Gyres and Mesoscales}
Figure \ref{fig:maps:NSH_streamlines_AVISO_NEMO} visualizes the flow from both AVISO satellite data and NEMO reanalysis model data (see Methods) from a single daily mean at scales larger than and smaller than $10^3~$km, termed ``gyre-scale'' and ``mesoscale,'' respectively. 
The color intensity illustrates the flow speed and is consistent with expectations that the large-scale flow is primarily composed of signals from the western boundary currents, while the small-scales are dominated by mesoscales fluctuations.
In the upper panels of Fig.~\ref{fig:maps:NSH_streamlines_AVISO_NEMO} we can see clearly several well-known oceanic gyre structures, including the Beaufort Gyre in the Arctic, the Weddell and the Ross gyres in the Southern Ocean near Antactica, the subtropical and subpolar gyres in the Atlantic and Pacific basins, and the ACC.
North Atlantic currents are also readily observable, including the North Atlantic Current, its northward fork to the Norwegian Atlantic Current, and the southward East Greenland Current. 
The agreement between AVISO and NEMO is remarkable.

It is worth emphasizing that the flows in Fig.~\ref{fig:maps:NSH_streamlines_AVISO_NEMO} are derived deterministically from a single daily mean of surface geostrophic velocity data without further temporal or statistical averaging.
Past approaches have used climatological multi-year averaging (e.g. \cite{aguiar2016seasonal,shi2021ocean}) or Empirical Orthogonal Function (EOF) analysis (e.g. \cite{Trenberth1975,DiLorenzo2008}), which is a statistical approach that requires averaging long time-series. 
Coarse-graining allows us to derive the dynamics governing the evolution of the flow in Fig.~\ref{fig:maps:NSH_streamlines_AVISO_NEMO} (e.g. \cite{aluie2018mapping}), which is not possible for EOF analysis, and to disentangle length-scales and time-scales independently and in a self-consistent manner to study interactions between different spatio-temporal scales that link large-scale forcing, the mesoscale eddy field, and the global-scale circulation. 
Such objective disentanglement of the oceanic circulation by coarse-graining opens the door for analysing the coupling of the ocean's mesoscales, on spatio-temporal scales of $\mathcal{O}$(100~km) and $\mathcal{O}$(30~days), to the global circulation ( $\mathcal{O}$($10^3~$km) and $\mathcal{O}$(10~years) ) and the climate system ( $\mathcal{O}$($10^4~$km) and $\mathcal{O}$(100~years) ).

\subsection*{Global Kinetic Energy Spectrum}

\begin{figure*}[ht]
    \centering
    \includegraphics[scale = 1]{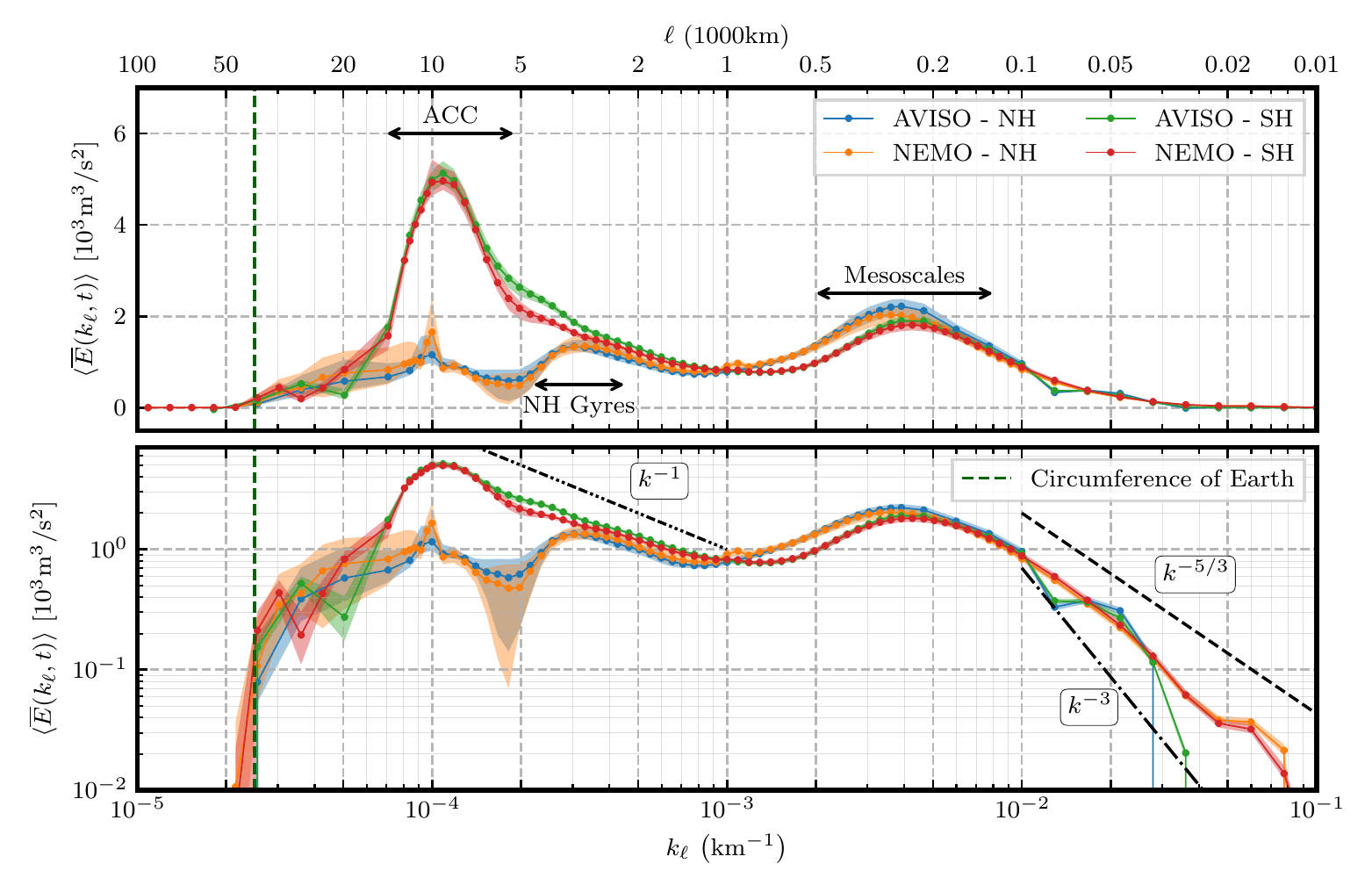}
    \caption{
        \textbf{Power Spectral Density}
        Filtering wavenumber spectra (see eq. \eqref{eq:filt_spect}) of surface geostrophic KE for the global extratropical ocean from AVISO satellite altimetry and NEMO model re-analysis. Northern and southern hemispheres (NH and SH, respectively) extend poleward of \( 15^\circ\). 
        Both panels show the same data, but using \textbf{[top]} lin-log and \textbf{[bottom]} log-log scales.
        Plots show the temporal mean, $\langle\cdot\rangle$, of $\olE(k_\ell,t)$ while
        envelopes show inter-quartile range (25th to 75th percentiles) of temporal variation.
        Data markers indicate length scales at which coarse-graining was performed.
        The vertical dashed green line at $40.0\times 10^3~$km indicates the equatorial circumference of the Earth.
        Dashed black lines provide a reference for \(-\sfrac{5}{3}\), \(-3\), and \(-1\) power-law slopes in the bottom panel.
    }
    \label{fig:spectra}
\end{figure*}

Figure \ref{fig:spectra} shows the filtering spectrum for both the northern and southern hemispheres as obtained from eq.\ \eqref{eq:filt_spect} using surface geostrophic velocity data from both satellite altimetry and a high-resolution model (see Methods). 
This is the first spectrum showing the oceanic energy distribution across such a wide range of scales, from planetary scales $\cO(10\times10^3)~$km down to $\cO(10)~$km. 

The top and bottom panels in Fig.~\ref{fig:spectra} plot the same spectrum in lin-log and log-log scale, respectively. 
The top panel highlights the prominent spectral peak due to the ACC, which is more than twice the mesoscale peak. 
The bottom panel highlights the power-law scaling over different $k_\ell$ bands. 

Note the zero energy content at scales larger than Earth's circumference and that energy also decreases precipitously when approaching the smallest scales resolved by each of the datasets, both of which are physical expectations.
It is not possible for simulation, satellite, or field data to capture all scales present in the natural ocean, which certainly has scales smaller than $100~$km. 
There is excellent agreement between satellite data and the higher resolution model data used here down to scales $\approx 100~$km, which indicates that all scales larger than $100~$km are well-resolved by both datasets, whereas smaller scales (10 -- 100 km) are reasonably resolved only in the model data.

\paragraph*{Antarctic Circumpolar Current and Oceanic Gyres}
Unlike previously reported KE wavenumber spectra using Fourier analysis on box regions (e.g. \cite{sasaki2014impact,Torresetal2018jgr,Qiuetal2018jpo}), some of which show a peak at mesoscales $O(100)~$km, our Fig.~\ref{fig:spectra} reveals that the largest spectral peak occurs at scales approximately $100$ times larger, at $\approx 10\times10^3~$km, and only in the southern hemisphere. 
Indeed, the ACC at latitude 50$^\circ$S has a geodesic diameter of $\approx8.9\times10^3~$km as measured from the South Pole. 
This can also be seen from the yellow color of the ACC in Fig.~\ref{fig:maps:NSH_streamlines_AVISO_NEMO}, highlighting its contribution to KE at large scales. 
Additional support that this spectral peak is due to the ACC can be found in Fig.~\ref{fig:zonally_avg_coarse_KE}, which plots the zonally (east-west) averaged KE as a function of latitude at various scales larger than $10^3~$km. 
We can see from Fig.~\ref{fig:zonally_avg_coarse_KE} that the dominant contribution is from latitudes [60$^\circ$S, 40$^\circ$S], which roughly corresponds with the ACC. 
We also see in Fig.~\ref{fig:zonally_avg_coarse_KE} a much weaker signal at latitudes [30$^\circ$N, 40$^\circ$N], which roughly aligns with the Gulf Stream and Kuroshio. 
Further corroborating our assertion, the spectral peak in the southern hemisphere seen in Fig.~\ref{fig:spectra} has no analogous peak in the northern hemisphere. 
Fig. \ref{fig:spectra} shows vestiges of a similar peak in the northern hemisphere, but this is arrested at a smaller amplitude and at smaller scales ($\approx 4\times10^3~$km) due to continental boundaries. 

\paragraph*{Gyre-scale Power Law}
Comparing the KE spectra from both hemispheres in Fig.~\ref{fig:spectra} at scales larger than \(10^3~\)km, we observe a range of scales that exhibit a $\sim k^{-1}$ power-law. 
This scaling has been predicted by \cite{MullerFrankignoul1981} (see also \cite{Wortham2014}) for baroclinic modes at scales larger than the barotropic deformation radius, but has not been observed until now. 
The barotropic deformation radius is about $2\times10^3$ km in the oceans \cite{Vallis17} and the ocean flow tends to be surface intensified as expected in a baroclinic flow \cite{wunsch1997vertical}. 
Thus, the $k^{-1}$ scaling observed in Fig.~\ref{fig:spectra} is consistent with \cite{MullerFrankignoul1981}.
Previous studies relying on Fourier analysis within box regions would have had difficulty detecting such scaling due to the box size artifacts. 
The $k^{-1}$ extends to larger scales and peaks at scales $\approx 4\times10^3~$km in the north, which is the average scale at which the flow starts feeling continental boundaries and gyres form. 
This can also be seen from the bright yellow color of the Gulf Stream and Kuroshio in Fig.~\ref{fig:maps:NSH_streamlines_AVISO_NEMO}, highlighting their contribution to KE at large scales. 
In the southern hemisphere, on the other hand, the $k^{-1}$ scaling extends up to the scale of the ACC, which encounters no continental barriers (in the latitudes of the Drake Passage) as it flows eastward around Antarctica.

\paragraph*{Mesoscale Eddies}
In Fig.~\ref{fig:spectra}, we find a second spectral peak between $100~$km and $500~$km, centered at ${\ell\approx 300~}$km, that is associated with the mesoscale flow. 
While we can see from Fig.~\ref{fig:spectra} that the mesoscales do not form the largest peak of the KE spectrum, their cumulative contribution between scales $100~$km and $500~$km greatly exceeds that of scales larger than $10^3~$km. 
This is because the mesoscale flow populates a wider range of wavenumbers compared to the gyre-scale flow (note the logarithmic x-axis). 
Indeed, integrating the energy spectrum in Fig.~\ref{fig:spectra} within the band $100~$km to $500~$km yields more than $\approx50\%$ of the total energy resolved by either satellites or the mesoscale eddying model in the extratropics. 
Coarse-graining allows us to determine this fraction of KE belonging to the mesoscales \emph{in the global ocean}. 
This is because integrating the filtering spectrum over all $k_\ell$ in Fig.~\ref{fig:spectra} yields the total KE (as resolved by the data), which was not possible in past studies using Fourier analysis in regional boxes.

The power-law spectral scaling at mesoscales and smaller scales has been the focus of many previous studies (e.g. \cite{Stammer1997ty,Uchida2017,Khatri2021, Qiuetal2018jpo}) using Fourier analysis within box regions. 
While this is not our focus here, we observe that the overall mesoscale spectral scaling lies between $k^{-5/3}$ and $k^{-3}$ in Fig.~\ref{fig:spectra}, consistent with previous studies \cite{Uchida2017,Xu2012}. 
Note that mesoscale power-law scaling is more clearly seen in smaller regions (e.g. Fig.~\ref{fig:Qiu_comparison} in Methods) as the mesoscale power-law and the corresponding wavenumber range change significantly depending on the geographical location (see Fig. 15 in \cite{Stammer1997ty}).

\paragraph*{ Characteristic Velocity and Energy Content within Key Scale Bands }
From the spectra in Fig.~\ref{fig:spectra}, we partition the energy \emph{conservatively} into four bands of interest: \({\ell\leq100~}\)km, 100~km to 500~km, 500~km to \(10^3\)~km, and \({\ell\geq10^3~}\)km such that the sum of their energy equals total KE. 
From KE within a scale band, $\mathrm{KE}_{\mathrm{band}}$, we can infer a characteristic root-mean-square (RMS) velocity, $u_\mathrm{rms}=\sqrt{2\times \mathrm{KE}_{\mathrm{band}}}$ at those scales.
The results are summarized in Table~\ref{table:ell_band_stats}.
The mesoscale band (100--500~km) has the highest RMS velocity, between 15 and 16 cm/s, and accounts for more than \(50\%\) of the total energy in the model data. 
Mesoscales are slightly more energetic in the NH than SH. 
Table~\ref{table:ell_band_stats} allows us to also infer a characteristic timescale, $\tau_\mathrm{meso} = \ell/u_\mathrm{rms} =\cO(25)~$days.
The RMS velocity decreases significantly for larger scales, with hemisphere-asymmetries becoming more prominent.
Within the ACC-containing band of \(\ell>10^3~\)km, the NH and SH RMS velocities are approximately 4.2 and 5.4 cm/s, respectively, and with an associated characteristic timescales, $\tau_\mathrm{gyre} =\ell/u_\mathrm{rms}=\cO(\mathrm{few})~$years.

\newcommand{\phz}{\hphantom{0}}
\setlength{\extrarowheight}{2pt}

\begin{table}[ht]
    \centering
    \begin{tabularx}{0.99\textwidth}{|c|c|c|c|c|}
    \cline{1-5}
    \multirow{4}*{\(\ell\)-band} & \multicolumn{4}{c|}{\(\sfrac{1}{12}^\circ\) NEMO} 
    \\ \cline{2-5}
         & \multicolumn{2}{c|}{RMS Vel.} & \multicolumn{2}{c|}{\% of }
    \\   & \multicolumn{2}{c|}{[cm/s]} & \multicolumn{2}{c|}{Total KE}
    \\\cline{2-5}
    & NH & SH & NH & SH
    \\\cline{1-5}\cline{1-5}
    \(\ell\leq100~\)km 
        & 13.15 & 13.29
        & 37.8  & 39.7
        \\\cline{1-5}
    100 to 500~km
        & 15.48 & 15.00
        & 53.2  & 50.2
        \\\cline{1-5}
    500 to 1000~km
        & \phz4.64 & \phz4.08
        & \phz4.7  & \phz3.7
        \\\cline{1-5}
    \(1000~\mathrm{km}\leq\ell\) 
        & \phz4.26 & \phz5.31 
        & \phz4.0  & \phz6.2 
        \\\cline{1-5} \cline{1-5}
    %
    %
    & \multicolumn{4}{c|}{\(\sfrac{1}{4}^\circ\) AVISO} 
    \\ \cline{1-5}
    \(\ell\leq100~\)km 
        & 11.21 & 11.24
        & 28.9  & 30.8
        \\\cline{1-5}
    100 to 500~km
        & 16.32 & 15.36
        & 61.9  & 57.7
        \\\cline{1-5}
    500 to 1000~km
        & \phz4.57 & \phz4.05
        & \phz4.9  & \phz4.0
        \\\cline{1-5}
    \(1000~\mathrm{km}\leq\ell\) 
        & \phz4.16 & \phz5.53 
        & \phz4.0  & \phz7.4 
        \\\cline{1-5}
    \end{tabularx}
    \caption{
        \textbf{Energy content of scale ranges}
        The RMS velocity in separate \(\ell\) bands for each hemisphere, as well as the percent of total KE contained within each \(\ell\) band, for both the \textbf{[upper half]} \(\sfrac{1}{12}^\circ\) NEMO and \textbf{[lower half]} \(\sfrac{1}{4}^\circ\) AVISO datasets.
        See Table \ref{table:ell_band_IQR} in Methods for uncertainty estimates.
    }
    \label{table:ell_band_stats}
\end{table}

\subsection*{Seasonality and Spectral Lag Time}

\begin{figure}[ht]
    \centering
    \includegraphics[scale=1]{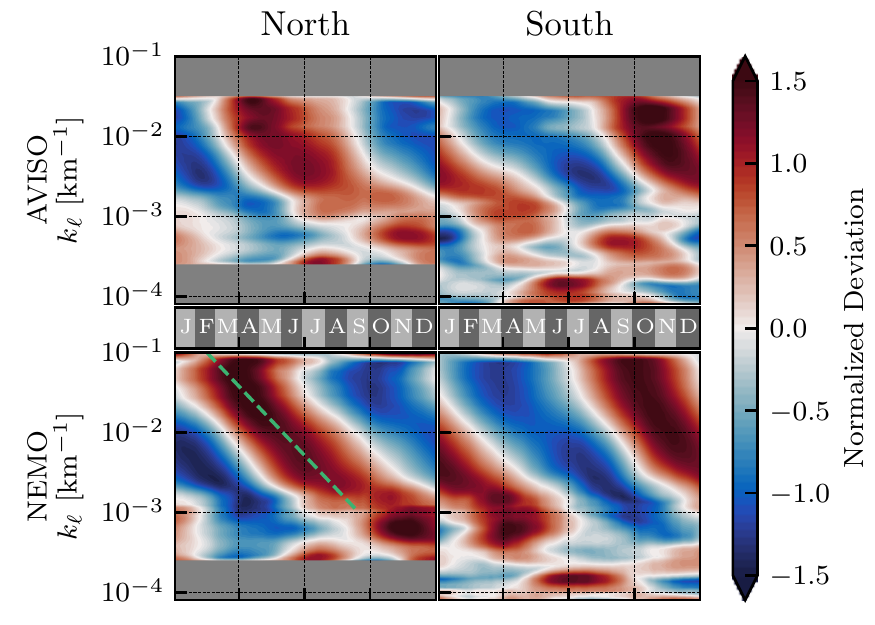}
    \caption{ \textbf{Seasonality}
        Normalized deviation (cf. Methods section) of the 60-day running average of surface kinetic energy in \textbf{[left]} NH  and \textbf{[right]} SH for both \textbf{[top]} satellite and \textbf{[bottom]} model datasets.
        Horizontal axis shows time binned into months. Vertical axes show  filtering wavenumber \(k_\ell = \ell^{-1}\).
        The green line in the bottom left panel shows a 100-fold scale increase over 8 months. 
    }
    \label{fig:Seasonality}
\end{figure}

Figure~\ref{fig:Seasonality} shows the seasonality in surface KE as a function of length-scale from both satellite and model data, which exhibit similar trends.
The most striking feature of Fig.~\ref{fig:Seasonality} is the approximately constant lag time between length-scales of the same ratio as they attain seasonal maxima (red) and minima (blue).
Going from 10~km up to $10^3~$km, length-scales that are $\times2$ larger experience a lag of \(\approx40~\)days in their seasonal cycle, such that in both hemispheres KE at $100~$km peaks in late spring while KE at $10^3~$km peaks in late summer. 
A detailed regression analysis is in the Methods section. 
These results agree with and extend previous analysis \cite{Qiuetal2014,Uchida2017,sasaki2017regionality} within regional boxes, which found that scales between 50--100~km have maximal KE in the spring while scales larger than $200~$km (but smaller than the box) tend to peak with a delay of one--two months. 
Possible explanations for the seasonal variation in KE at different scales include the increased eddy-killing from winter's high winds \cite{Raietal2021}, and an inverse energy cascade from the submesocales which energizes mesoscales in spring months \cite{Uchida2017,sasaki2017regionality}.
While Fig.~\ref{fig:Seasonality} is suggestive of an inverse cascade, in which seasonal variations propagate up-scale at the rate we observe, it alone is not sufficient evidence (see \cite{Zhao2022}) and a direct measurement of the cascade as in \cite{aluie2018mapping} is required but is beyond our scope here. 

\paragraph*{ Gyre-scales }
At gyre-scales the surface flow is influenced directly by continental boundaries, wind, and buoyancy forcing. 
Indeed, at scales $>10^3~$km in Fig.~\ref{fig:Seasonality}, there is a noticeable break in the seasonal trends we discussed in the previous paragraph. 
In the SH, where the ACC is not impeded by continents, we see from Fig.~\ref{fig:Seasonality} a pronounced winter peak at $\approx10\times10^3~$km, which correlates with maximal wind forcing \cite{Raietal2021}. 
Scales between $10^3~$km and $3\times10^3~$km in both hemispheres peak in autumn, consistent with previous analysis showing an autumn maximum in the surface flow of western boundary currents \cite{WangKoblinsky1995,WangKoblinsky1996,Kelly+1999} due to the upper ocean seasonal heating cycle. 
At NH scales larger than $5\times10^3~$km, the KE is too small to be meaningful (Fig.~\ref{fig:spectra}).

\subsection*{Conclusion and Outlook}
Our spectral characterization of the ocean's surface velocity over the entire range of length-scales resolved by satellites and models revealed the ACC as the spectral peak of the extratropical ocean at $\approx10\times10^3~$km, with gyres in the northern hemisphere yielding a smaller peak that is arrested at $\approx4\times10^3~$km due to continental boundaries. 
By partitioning kinetic energy across length-scales in a manner that conserves energy and covers the global ocean, we showed that length-scales ${\ell<500~}$km make an overwhelming contribution to surface kinetic energy due to populating a wide range of wavenumbers, despite not forming the most prominent spectral peak. 
Based on prior characterization of ocean energy \cite{FerrariWunsch09}, we reason that these length scales are dominated by mesoscale features such as geostrophic turbulence, boundary currents, and fronts.
Our analysis also revealed a characteristic lag time, with length-scales that are twice as large experiencing a lag of \(\approx40~\)days in their seasonal cycle.

The expanded spectral analysis spurs new questions and lines of inquiry. 
We hope future investigations will shed light on the dynamic coupling between the spectral peaks at the gyre- and meso-scales, determine if the $k^{-1}$ slope between the two peaks is indeed due to baroclinic modes \cite{MullerFrankignoul1981,Wortham2014}, and whether the characteristic spectral lag-time is caused by an inverse cascade. 

\bibliographystyle{plain}
\bibliography{Ocean}

\begin{enumerate}
\item[] {\bf Acknowledgements} We thank D. Balwada, M. Jansen, and S. Rai for valuable discussions and comments. We also thank Bo Qiu for kindly sharing the data used in Fig. \ref{fig:Qiu_comparison}.
This research was funded by US NASA grant 80NSSC18K0772 and NSF grant OCE-2123496. HA was also supported by US DOE grants DE-SC0014318, DE-SC0020229, DE-SC0019329, NSF grant PHY-2020249, and US NNSA grants DE-NA0003856, DE-NA0003914. Computing time was provided by NERSC under Contract No. DE-AC02-05CH11231 and NASA's HEC Program through NCCS at Goddard Space Flight Center. This work was also supported by the European Research Council (ERC) under the European Union’s Horizon 2020 research and innovation programme (Grant Agreement No. 882340).
    \item[] {\bf Competing Interests} The authors declare that they have no
            competing financial interests.
        \item[] {\bf Correspondence} Correspondence and requests for materials
            should be addressed to H.A.\\Email: hussein@rochester.edu
\end{enumerate}

\twocolumn
\appendix
 
\makeatletter
\renewcommand{\thefigure}{E\@arabic\c@figure}
\renewcommand{\thetable}{E\@arabic\c@table}
\makeatother
\setcounter{figure}{0} 
\setcounter{table}{0} 
 \renewcommand{\theequation}{M-\arabic{equation}}
\setcounter{equation}{0}  
\renewcommand{\thepage}{{\it Methods -- \arabic{page}}} \setcounter{page}{1}

\section*{Methods \lb{sec:Methods}}

\subsection* {Description of datasets}
For the geostrophic ocean surface currents, we use Level 4 (L4) post-processed dataset of daily-averaged geostrophic velocity on a  $\sfrac{1}{4}^{\circ}$ grid and spanning  January 2010 to October 2018 (except for the seasonality analysis, where we use 2012-2016). The data is obtained from the AVISO$+$ analysis of multi-mission satellite altimetry measurements for sea surface height (SSH)~\citeSupp{pujol2016Supp}. The product identifier of the AVISO dataset used in this work is ``\sloppy{{\small SEALEVEL\_GLO\_PHY\_L4\_REP\_OBSERVATIONS\_008\_047}}''.

We also analyze 1-day averaged surface SSH-derived currents from the NEMO numerical modeling framework, which is coupled to the Met Office Unified Model atmosphere component, and the Los Alamos sea ice model (CICE). 
The NEMO dataset consists of weakly coupled ocean-atmosphere data assimilation and forecast system,
which is used to provide 10 days of 3D global ocean forecasts on a $\sfrac{1}{12}^\circ$ grid. 
We use daily-averaged  data that spans four years, from 2015 to 2018. 
More details about the coupled data assimilation system used for the production of the NEMO dataset can be found in~\citeSupp{gmd-4-223-2011Supp,lea2015assessingSupp}. 
The specific product identifier of the NEMO dataset used here is ``\sloppy{{\small GLOBAL\_REANALYSIS\_PHY\_001\_030}}''.

\subsection* {Coarse-graining on the sphere}
For a field $\phi(\bx)$,
a ``coarse-grained'' or (low-pass) filtered field, which contains only length-scales larger than $\ell$, is defined as
\be
\OL \phi_\ell(\bx) = G_\ell * \phi,
\lb{eqDefCoarseGrain}
\ee
where $*$, in the context of this work, is a convolution on the sphere as shown in \citeSupp{aluie2019convolutionsSupp} and $G_\ell(\br)$ is a normalized kernel (or window function) so that $\int d^2\br ~G_\ell(\br)=1$. 
Operation (\ref{eqDefCoarseGrain}) may be interpreted as a local space average over a region of diameter $\ell$ centered at point $\bx$, analogous to a moving time average.
The kernel $G_\ell$ that we use here is essentially a graded top-hat kernel:
\begin{equation}
    G_{\ell}(\bx) = \frac{A}{2}\left( 1 - \tanh\left( 10\left (\frac{\gamma(\bx)}{\ell/2}-1 \right ) \right) \right).
    \label{eq:tanh-filter}
\end{equation}
We use geodesic distance, $\gamma(\bx)$, between any location $\bx=(\lambda,\phi)$ on Earth's surface relative to location $(\lambda_0,\phi_0)$ where coarse-graining is being performed, which we calculate using
\begin{equation}
    \gamma(\bx) = R_{\text{\tiny{E}}}\arccos\Big[\sin(\phi)\sin(\phi_0)+\cos(\phi)\cos(\phi_0)\cos(\lambda-\lambda_0)\Big],
    \label{eq:HaversineDistance}
\end{equation}
with $R_{\text{\tiny{E}}}=6371~$km for Earth's radius. In eq. \eqref{eq:tanh-filter}, $A$ is a normalization factor, evaluated numerically, to ensure $G_\ell$ area integrates to unity. 
A convolution with $G_\ell$ in equation \eqref{eq:tanh-filter} is a spatial analogue to an  $\ell$-day running time-average. 

The above formalism holds for coarse-graining scalar fields. 
To coarse-grain a vector field on a sphere generally requires more work \citeSupp{aluie2019convolutionsSupp}, particularly for vector fields that need not be toroidal (2D non-divergent) or potential (2D-irrotational).
However, as this work focuses on SSH-derived 2D non-divergent velocity fields, these concerns do not apply here. 
More details can be found in \citeSupp{buzzicotti2021Supp}.

\subsubsection* {Comparing coarse-graining to Fourier analysis}
It is common to quantify the spectral distribution of ocean kinetic energy via Fourier transforms computed either along transects or within regions; e.g.,  \citeSupp{FuSmith1996Supp,Chenetal2015Supp,Rochaetal2016Supp,khatri2018surfaceSupp,ORourke_etal2018Supp,CalliesWu2019Supp}. This approach has rendered great insights into the length scales of oceanic motion and the cascade of energy through these scales \citeSupp{ScottWang05Supp,ScottArbic07Supp,arbic2012nonlinearSupp,Arbicetal13Supp,arbic2014geostrophicSupp}. However, it has notable limitations for the ocean where the spatial domain is generally not periodic, thus necessitating adjustments to the data (e.g., by tapering) before applying Fourier transforms. Methods to produce an artifically periodic dataset can introduce spurious gradients, length-scales, and flow features not present in the original data \citeSupp{sadek2018extractingSupp}. 
A related limitation concerns the chosen region size, with this size introducing an artificial upper length scale cutoff. In this manner, no scales are included that are larger than the region size even if larger structures exist in the ocean. Furthermore, the data is typically assumed to lie on a  flat tangent plane to enable the use of Cartesian coordinates. However, if the region becomes large enough to sample the earth's curvature, then that puts into question the use of the familiar Cartesian Fourier analysis of sines and cosines. The use of spherical harmonics, common for the atmosphere, is not suitable for the ocean, again since the ocean boundaries are complex. These limitations mean that in practice, Fourier methods are only suited for open ocean regions away from boundaries, and over a rather limited regional size.

\begin{figure}[ht]
    \centering
    \includegraphics[scale=1]{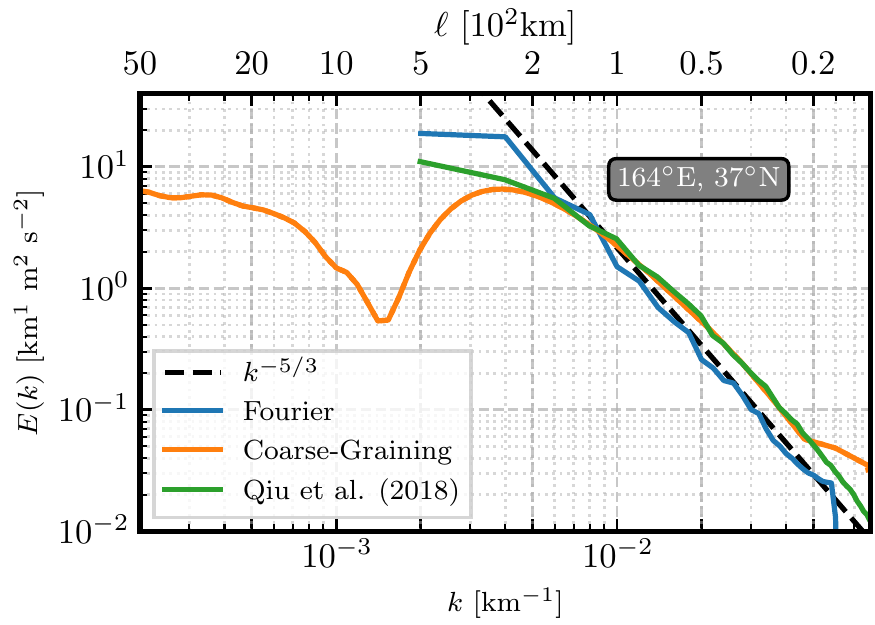}
    \caption{
        \textbf{Energy Spectra.}
         Comparison of the filtering spectrum (orange), traditional Fourier spectrum (blue), and the Fourier spectrum from [Supp14] (green, data kindly provided by authors of [Supp14]) for the \({5^\circ\times5^\circ}\) box region centred at \(164^\circ\mathrm{E},~37^\circ\mathrm{N}\) (roughly the Kuroshio extension). The dashed black line provides a reference for a \(-\sfrac{5}{3}\) slope. Note that the \(\ell\) axis (top) is in \(100\)~km.
        }
    \label{fig:Qiu_comparison}
\end{figure}

As a demonstration of both the validity and advantages of coarse-graining for energy spectra, consider Figure \ref{fig:Qiu_comparison}.
This figure reproduces the energy spectrum from Figure 3 of \citeSupp{Qiuetal2018jpoSupp}, and includes both the coarse-graining, and traditional Fourier energy spectra measured from the \(\sfrac{1}{12}^\circ\) NEMO dataset. Spectra are calculated for the \(5^\circ\times5^\circ\) box centred at \(164^\circ\mathrm{E},~37^\circ\mathrm{N}\), which corresponds to the Kuroshio current.

For length scales \({\lesssim200~}\)km, the three spectra generally agree very well, and all produce close to a \(-\sfrac{5}{3}\) spectrum. 
\citeSupp{Qiuetal2018jpoSupp} used a higher resolution dataset, and so the spectra diverge for very small scales.
However, coarse-graining does not require tapering, and so the spectrum at scales \({\gtrsim200~}\)km are not contaminated by the shape of the tapering window.
As a result, coarse-graining is able to detect that the spectrum for this region peaks at \({\sim250~}\)km, with a minimum near \(600~\)km.

\subsubsection* {ACC as the spectral peak}

\begin{figure}[ht]
    \includegraphics[scale=1]{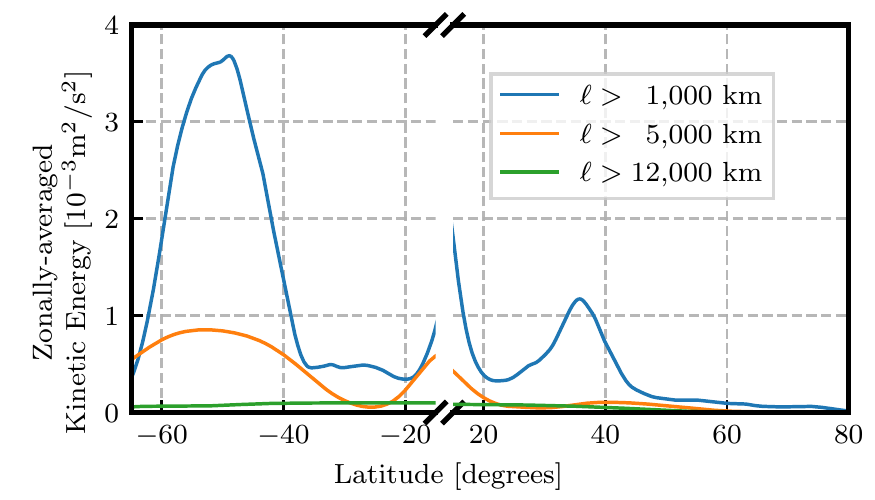}
    \caption{
        \textbf{Energy by Latitude} 
        Time- and zonally-averaged kinetic energy computed from AVISO as a function of latitude for a selection of filter scales (see legend). 
        Note that the latitude axis is broken to exclude the band [$15^\circ$S, $15^\circ$N].
    }
    \label{fig:zonally_avg_coarse_KE}
\end{figure}

In Figure~\ref{fig:zonally_avg_coarse_KE} we provide a visualization of the zonally-averaged kinetic energy for selected filtering scales. 
Scales larger than $10^3$~km have a dominant contribution from latitudes [60$^\circ$S, 40$^\circ$S], roughly corresponding with the ACC, and another contribution over [30$^\circ$N, 40$^\circ$N], roughly corresponding to the NH boundary currents.
Scales larger than $5\times10^3$~km continue to show a clear ACC signal, with no NH signal since this filter scale is just beyond the NH gyre spectral peak. 
Finally, scales larger than $12\times10^3~$km have no distinct ACC signal, showing that the ACC has been fully removed by this scale.
Combined, these provide further support for our claim that the \(10\times10^3~\)km spectral peak corresponds to the ACC.

\subsubsection*{Land Treatment}

When coarse-graining near land, it is necessary to have a methodology for incorporating land into the filtering kernel (c.f. \citeSupp{buzzicotti2021Supp} for more in-depth discussion of land treatments).
In the work presented here, we make the choice of treating land as zero velocity water.
Since coarse-graining is essentially a `blurring' (analogized with taking off ones glasses to have a blurrier picture), the land-water division itself also become less well-defined, and so treating land as zero-velocity water is both conceptually consistent and aligns with no-flow boundary conditions.
Additionally, this land treatment allows for a `fixed' (or homogeneous) filtering kernel at all points in space, and as a result allows for commutativity with derivatives (i.e. divergence-free flows remain divergence-free after coarse-graining) \citeSupp{aluie2019convolutionsSupp}.
Note, however, that only the true water area is used as the denominator when computing area averages (e.g. the NH area-average energy is the coarse energy summed over \textit{all} NH cells, including land, divide by the \textit{water-area} of NH).
The ocean areas used for the area-averaging are \(\approx104\times10^6~\mathrm{ km}^2\) for NH and \(\approx155\times10^6~\mathrm{ km}^2\) for SH.

\paragraph{Deforming kernel approach}
An alternative choice is to \textit{deform} the kernel around land, so that only water cells are included, at the cost of losing the homogeneity of the kernel.
The benefit to this approach is that it does not require conceptually treating land as zero velocity water.
However, it has the significant drawback that coarse-graining no longer commutes with differentiation and, as a result, does not necessarily preserve flow properties such as being divergence-free.
Additionally, a kernel that is inhomogeneous (i.e. changes shape depending on geographic location) does not necessarily conserve domain averages, including the kinetic energy of the flow, and has the potential to both increase or decrease the domain average. 
More details are provided in \citeSupp{buzzicotti2021Supp}.

\begin{figure}[ht]
    \centering
    \includegraphics{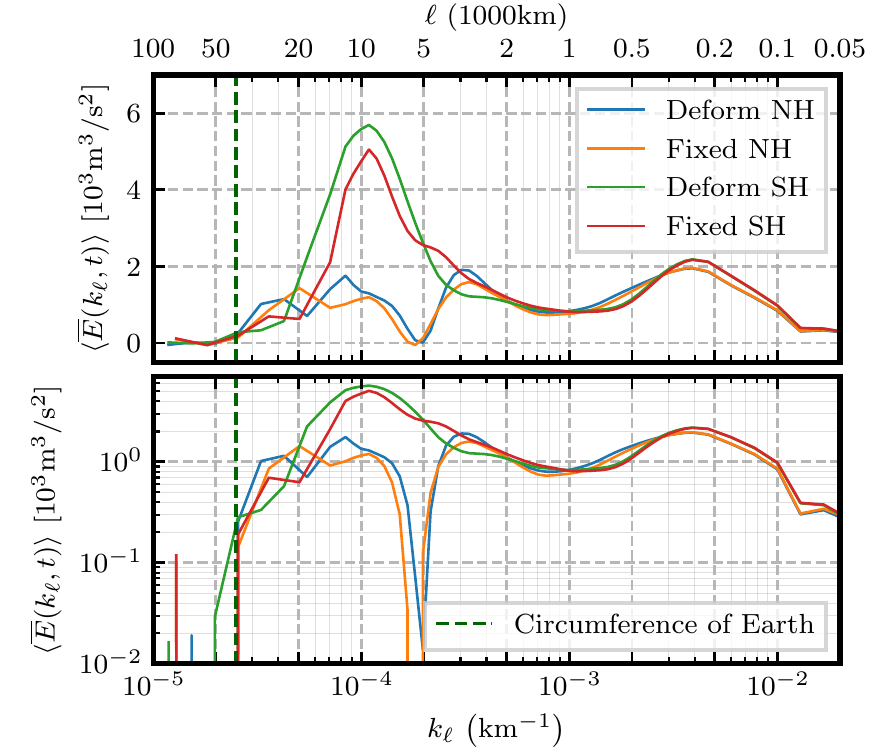}
    \caption{
        Filtering spectra, analogous to Fig.~\ref{fig:spectra}, using both deforming and fixed kernels on the AVISO dataset for a single day (02 Jan 2015).
    }
    \label{fig:deforming_kernel}
\end{figure}

\paragraph{Comparing land treatments}
Figure \ref{fig:deforming_kernel} presents the energy spectra, similar to Fig.~\ref{fig:spectra} in the main text, using both deforming and fixed kernels for the single day 02 Jan 2015.
The deforming-kernel spectra agree remarkably well with the non-deforming (fixed) kernel spectra, in that they present the mesoscales, ACC, and gyre peaks in similar locations. There are some quantitative differences, such as
the deforming kernel SH spectra presents a slightly broader and higher-magnitude ACC peak.

\subsubsection*{ Isolating Hemisphere Spectra }

In this work, we are primarily concerned with the extra-tropical latitudes: \([90^\circ\mathrm{S},15^\circ\mathrm{S}]\) and \([15^\circ\mathrm{N},90^\circ\mathrm{N}]\).
However, at very large length scales information from the equatorial band and opposing hemisphere can become introduced through the expanding filter kernel.
To resolve this issue, we use a `reflected hemispheres' approach, wherein one hemisphere is reflected and copied onto the other hemisphere, essentially producing a world with two north, or two south hemispheres.
It is worth noting that reflected hemispheres and equatorial masking would not be necessary in a context where ageostrophic velocities are also considered and a global power spectrum is desired.
They are used here because we wish to disentangle the power spectra of the extra-tropical hemispheres separately.

\begin{figure}[ht]
    \centering
    \includegraphics[scale=1]{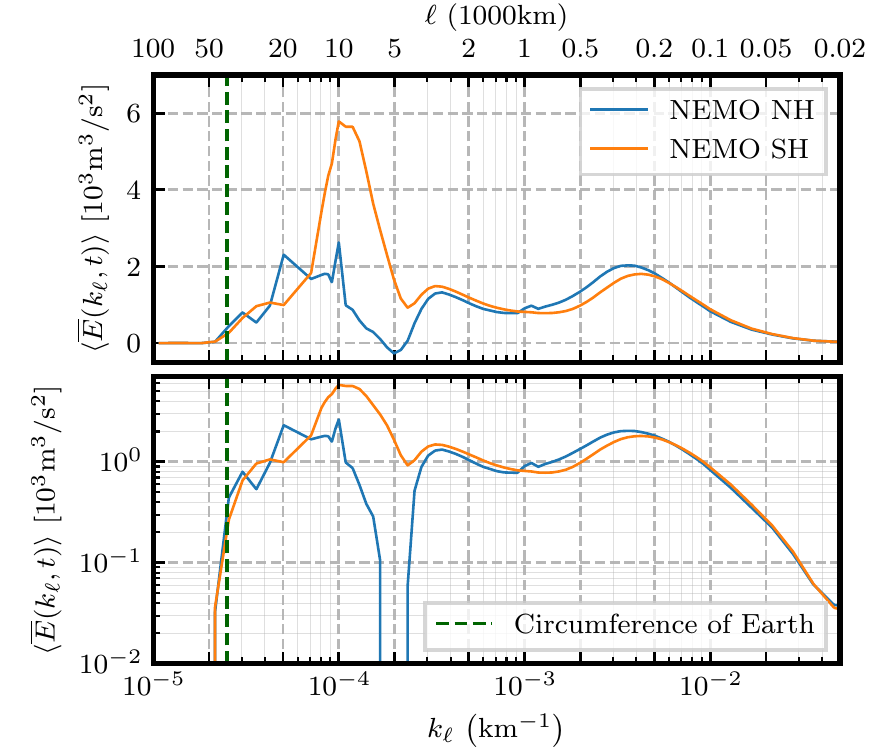}
    \caption{
        Same layout as Fig.~\ref{fig:spectra}, but without hemisphere reflections.
        That is, NH and SH spectra are extracted from a single global coarse-graining.
        Note that the NH spectra has a large-scale peak near \(\ell=20\times10^3~\)km, which is not observed using reflected hemispheres. 
        Produced using the same four-year period of NEMO data, sub-sampled to every fourth day.
    }
    \label{appendix:reflected_hemispheres}
\end{figure}

Figure \ref{appendix:reflected_hemispheres} shows the filtering spectra from NEMO without relying on hemisphere reflection, and is to be compared to Fig.~\ref{fig:spectra} in the main text. 
The two are in qualitative agreement, with an ACC peak in the SH and mesoscale peaks in both hemispheres. 
Unsurprisingly, the spectra only deviate for very large filtering scales, where an increasing amount of extra-hemisphere information is captured by the large kernels.
Specifically, the NH spectra has a third peak at scales \(\ell>10\times10^3~\)km that is not present when using reflected hemispheres.
This very large-scale peak is a result of the NH kernels capturing the ACC.
It is worth noting, however, that the main ACC peak is still present in the SH spectra, as is the NH gyre peak at approximately \(\ell=3\times10^3~\)km.

\subsection*{Seasonality}
\label{sec:z_score}
A 60-day running mean is applied to remove higher frequencies and allow us to better consider the longer-time trend, and individual years are averaged onto a `typical' year for the purpose of comparison.
Seasonality results using the 5 years spanning 2012-2016 for AVISO, and the 4 years spanning 2015-2018 for NEMO.

A useful statistical method for comparing signals is to compare the normalized deviation, or z-scores, of the signal.
For a set of points \(\{x_i \,|\, i=1\ldots N\}\), each point \(x_i\) is transformed into a corresponding z-score \(z_i\) via \(z_i = (x_i - \mu_x)/\sigma_x\), where \(\mu_x\) and \(\sigma_x\) are the mean and standard deviation of the \(x_i\).
As a result, data points that are larger than the mean produce a positive z-score, while those smaller than the mean produce a negative z-score.
Note that the normalized deviation (z-scores) in Fig.~\ref{fig:Seasonality} are computed independently for each $k_\ell$, and so comparing magnitudes between scales is non-trivial.

\subsubsection*{ Regression Analysis of Phase Shift }

\begin{figure}[ht]
    \centering
    \includegraphics[scale=1]{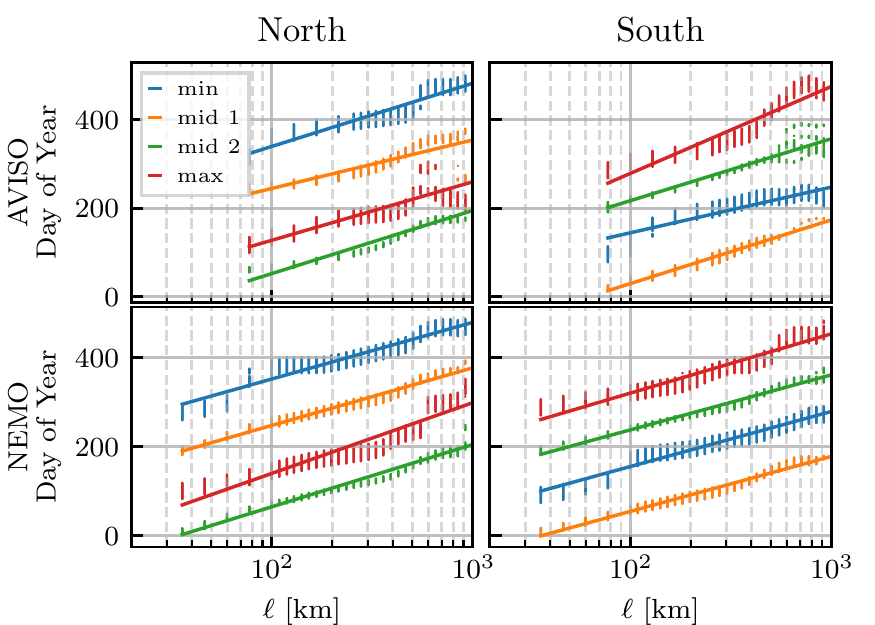}
    \caption{
        Extracted data points and regression fits for the seasonal drift exhibited in Fig.~\ref{fig:Seasonality}.
        Panel layout ordering is identical to Fig.~\ref{fig:Seasonality}.
        Dots ( which appear as vertical bars due to density ) show the extracted \(k_{\ell}\)-time points, and the lines show the corresponding regression fit.
        In the legend, `min' means the lowest 10\% z-score, `mid' the two middle-most 10\% groups, and `max' the highest 10\%.
        }
    \label{fig:append:seasonality_trends}
\end{figure}

Fig.~\ref{fig:Seasonality} presents a clear phase shift in the seasonal cycle as a function of length-scale.
In order to quantify the phase shift, we need to first extract a meaningful set of \((k_\ell,\mathrm{time})\) points.
To that end, we extract, for each \(k_\ell\), the i) times corresponding to the lowest 10\%, ii) middle-most 10\%, and iii) highest 10\% of the normalized deviations presented in Fig.~\ref{fig:Seasonality}.
Heuristically, this is extracting the \((k_\ell,\mathrm{time})\)-coordinates for the line of darkest red, darkest blue, and the two white lines, results in a total of four regression sets.
Where necessary, periodic phase adjustments are applied to maintain monotonicity in time, and the \(k_\ell\) grid is truncated to focus on regimes with a clear linear trend.
The extracted data points are shown as the dots / vertical bars in Figure \ref{fig:append:seasonality_trends}, along with their corresponding regression fits.

\begin{figure}[ht]
    \centering
    \includegraphics[scale=1]{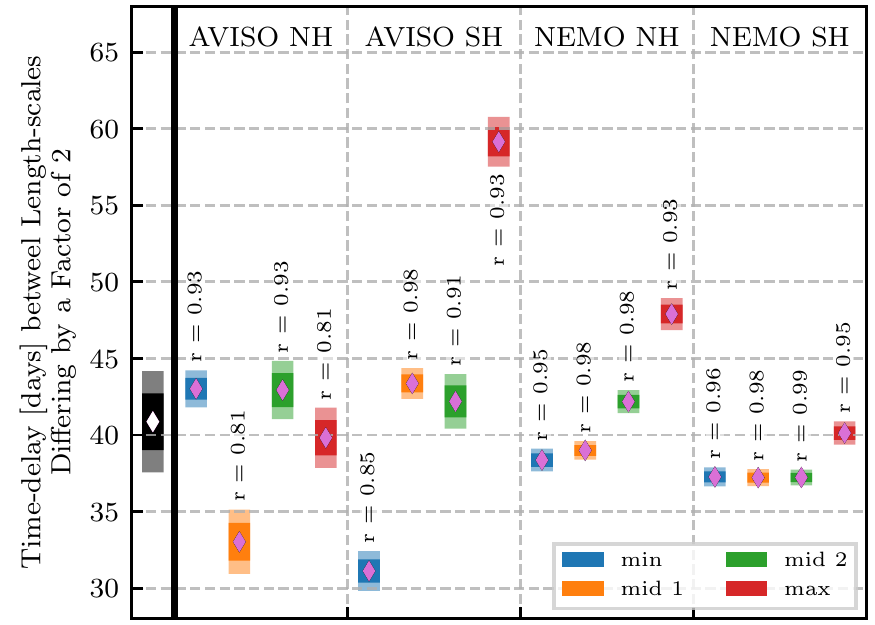}
    \caption{ 
        Regression slope estimates and confidence intervals for each of the datasets shown in Fig.~\ref{fig:append:seasonality_trends}.
        Vertical dashed lines separate the data sources, with the text along the top indicating the source data and hemisphere.
        Diamonds indicate the regression slope estimate, dark envelopes the 75\% confidence interval, and light envelopes the 95\% confidence interval.
        The r-value for each regression fit is printed alongside the corresponding slope distribution.
        The left-most illustration, separated by a thick black line, presents the mean (diamond) and confident intervals (envelopes) across the 16 regression analysis.
        In the legend, `min' means the lowest 10\% z-score, `mid' the two middle-most 10\% groups, and `max' the highest 10\%.
    }
    \label{fig:append:seasonality_timescale}
\end{figure}

Figure \ref{fig:append:seasonality_timescale} presents the linear regression slope analysis for the data shown in Fig.~\ref{fig:append:seasonality_trends}.
The different regression analyses generally agree well, with 12 of the 16 regression sets indicating at 35--45~day time-lag per octave of spatial scale.
From this analysis, we conclude that length-scales that differ by a factor of two (i.e. \(\ell_1/\ell_2=2\)) have seasonal cycles that are off-set by \(41\pm3\) days.
Scales that differ a decade (\(\ell_1/\ell_2=10\)) would correspondingly have a phase shift of \(136\pm10~\)days, or roughly 4.5 months.

\subsection*{ Uncertainty estimates of \(\ell\)-band Values }

Table \ref{table:ell_band_stats} presented median values of the RMS velocity and percentage of total KE contained within various \(\ell\)-bands.
Supplemental Table \ref{table:ell_band_IQR} presents the interquartile range (25$^{\mathrm{th}}$ to 75$^{\mathrm{th}}$ percentiles) to provide an estimate for the sensitivity of those values.

\begin{table*}[ht]
    \centering
    \begin{tabularx}{0.60\textwidth}{|c|c|c|c|c|}
    \cline{1-5}
    \multirow{4}*{\(\ell\)-band} & \multicolumn{4}{c|}{\(\sfrac{1}{12}^\circ\) NEMO} 
    \\ \cline{2-5}
         & \multicolumn{2}{c|}{RMS Vel.} & \multicolumn{2}{c|}{\% of }
    \\   & \multicolumn{2}{c|}{[cm/s]} & \multicolumn{2}{c|}{Total KE}
    \\\cline{2-5}
    & NH & SH & NH & SH
    \\\cline{1-5}\cline{1-5}
    \(\ell\leq100~\)km 
        & 12.77 to 13.56 & 12.98 to 13.78
        & 36.3  to 40.5  & 38.2  to 41.0
        \\\cline{1-5}
    100 to 500~km
        & 14.97 to 16.16 & 14.57 to 15.46
        & 50.8  to 54.6  & 49.0  to 51.5
        \\\cline{1-5}
    500 to 1000~km
        & \phz4.47 to \phz4.79 & \phz3.99 to \phz4.16
        & \phz4.3  to \phz5.1  & \phz3.4  to \phz3.9
        \\\cline{1-5}
    \(1000~\mathrm{km}\leq\ell\) 
        & \phz4.11 to \phz4.43 & \phz5.27 to \phz5.36
        & \phz3.6  to \phz4.4  & \phz5.9  to \phz6.6
        \\\cline{1-5} \cline{1-5}
    & \multicolumn{4}{c|}{ } 
    \\
    & \multicolumn{4}{c|}{\(\sfrac{1}{4}^\circ\) AVISO} 
    \\ \cline{1-5}
    \(\ell\leq100~\)km 
        & 10.89 to 11.55 & 11.04 to 11.57
        & 28.3  to 30.2  & 30.1  to 31.7
        \\\cline{1-5}
    100 to 500~km
        & 15.62 to 16.83 & 15.02 to 15.79
        & 60.7  to 62.8  & 56.7  to 58.5
        \\\cline{1-5}
    500 to 1000~km
        & \phz4.46 to \phz4.68 & \phz3.98 to \phz4.12
        & \phz4.6  to \phz5.2  & \phz3.8  to \phz4.2
        \\\cline{1-5}
    \(1000~\mathrm{km}\leq\ell\) 
        & \phz4.06 to \phz4.26 & \phz5.47 to \phz5.57
        & \phz3.7  to \phz4.4  & \phz7.0  to \phz7.7 
        \\\cline{1-5}
    \end{tabularx}
    \caption{
        \textbf{Energy content of scale ranges}
        Interquartile range of RMS velocity values (left half) and percent of total kinetic energy in separate \(\ell\) bands in each hemisphere, for both the \textbf{[upper half]} \(\sfrac{1}{12}^\circ\) NEMO and \textbf{[lower half]} \(\sfrac{1}{4}^\circ\) AVISO datasets.
    }
    \label{table:ell_band_IQR}
\end{table*}

\section*{Data availability}
All other data that support the findings of this study are available from the corresponding author on reasonable request.

\bibliographystyleSupp{plain}
\bibliographySupp{OceanSupp}

\end{document}